\documentclass[runningheads]{llncs}
\usepackage{graphicx}
\usepackage{url}
\usepackage{marvosym}
\usepackage{multirow}
\usepackage{amsfonts}
\usepackage{algorithm}
\usepackage{algpseudocode}
\usepackage{amsmath}

\usepackage{todonotes}

\begin{document}
\title{Localizing the Recurrent Laryngeal Nerve via Ultrasound with a Bayesian Shape Framework}
\titlerunning{Localizing Recurrent Laryngeal Nerve with Bayesian Shape Framework}
%
\author{Haoran Dou\inst{3,4}\thanks{Haoran Dou and Luyi Han contributed equally to this work.}
\and
Luyi Han\inst{5,6\star}
\and
Yushuang He\inst{7}
\and
Jun Xu\inst{2}
\and
Nishant Ravikumar\inst{3,4}
\and
Ritse Mann\inst{5, 6}
\and
Alejandro F. Frangi\inst{3, 4, 8, 9, 10}
\and
Pew-Thian Yap\inst{11}
\and
Yunzhi Huang\inst{1}\textsuperscript{(\Letter)}}
\authorrunning{H. Dou et al.}
%
\institute{Institute for AI in Medicine, School of Automation, Nanjing University of Information Science and Technology, Nanjing, China \and
Institute for AI in Medicine, School of Artificial Intelligence, Nanjing University of Information Science and Technology, Nanjing, China \and
Centre for Computational Imaging and Simulation Technologies in Biomedicine (CISTIB), School of Computing, University of Leeds, Leeds, UK \and
Biomedical Imaging Department, Leeds Institute for Cardiovascular and Metabolic Medicine (LICAMM), School of Medicine, University of Leeds, Leeds, UK 
\and
Department of Radiology and Nuclear Medicine, Radboud University Medical Centre, Nijmegen, The Netherlands
\and
Department of Radiology, Netherlands Cancer Institute (NKI), Amsterdam, The Netherlands 
\and
West China Hospital of Sichuan University, Chengdu, China \and
Department of Cardiovascular Sciences, KU Leuven, Leuven, Belgium
\and
Department of Electrical Engineering, KU Leuven, Leuven, Belgium
\and
Alan Turing Institute, London, UK
\and 
Department of Radiology and Biomedical Research Imaging Center (BRIC), University of North Carolina, Chapel Hill, USA
\\
\email{\{yunzhi.huang.scu\}@gmail.com}}
\maketitle              
\begin{abstract}
Tumor infiltration of the recurrent laryngeal nerve (RLN) is a contraindication for robotic thyroidectomy and can be difficult to detect via standard laryngoscopy. Ultrasound (US) is a viable alternative for RLN detection due to its safety and ability to provide real-time feedback. However, the tininess of the RLN, with a diameter typically less than 3\,mm, poses significant challenges to the accurate localization of the RLN. In this work, we propose a knowledge-driven framework for RLN localization, mimicking the standard approach surgeons take to identify the RLN according to its surrounding organs. We construct a prior anatomical model based on the inherent relative spatial relationships between organs. Through Bayesian shape alignment (BSA), we obtain the candidate coordinates of the center of a region of interest (ROI) that encloses the RLN. The ROI allows a decreased field of view for determining the refined centroid of the RLN using a dual-path identification network, based on multi-scale semantic information.
Experimental results indicate that the proposed method achieves superior hit rates and substantially smaller distance errors compared with state-of-the-art methods. 

\keywords{Bayesian Shape Alignment \and Recurrent Laryngeal Nerve \and Localization in Ultrasound.}
\end{abstract}
\section{Introduction}
\label{introduction}
Robotic thyroidectomy safely removes low-risk tumors and is preferred by patients who want a scarless operation with minimal invasiveness~\cite{JandeeLee2013RoboticSF,KaiPunWong2013EndoscopicTA}. A complete pre-operative assessment of the surroundings of the thyroid is critical for accurate surgical planning to prevent unnecessary harm to the internal jugular vein or the recurrent laryngeal nerve (RLN). Currently, laryngoscopy is the only method surgeons can rely on to detect whether the RLN is tumor-infiltrated. However, laryngoscopy determines the RLN indirectly by assessing the activity of the vocal cords~\cite{SujanaSChandrasekhar2013ClinicalPG}. This approach is therefore highly inaccurate and can only distinguish RLN abnormality in about $1$--$3\%$ patients~\cite{GianlorenzoDionigi2010PostoperativeLI}. Recent clinical trials have turned to ultrasound (US) as an alternative method for pre-operative inspection due to its safety and ability to provide real-time feedback~\cite{YushuangHe2021PreoperativeVU}. 

\begin{figure*}
    \centering
    \includegraphics[width=0.7\linewidth]{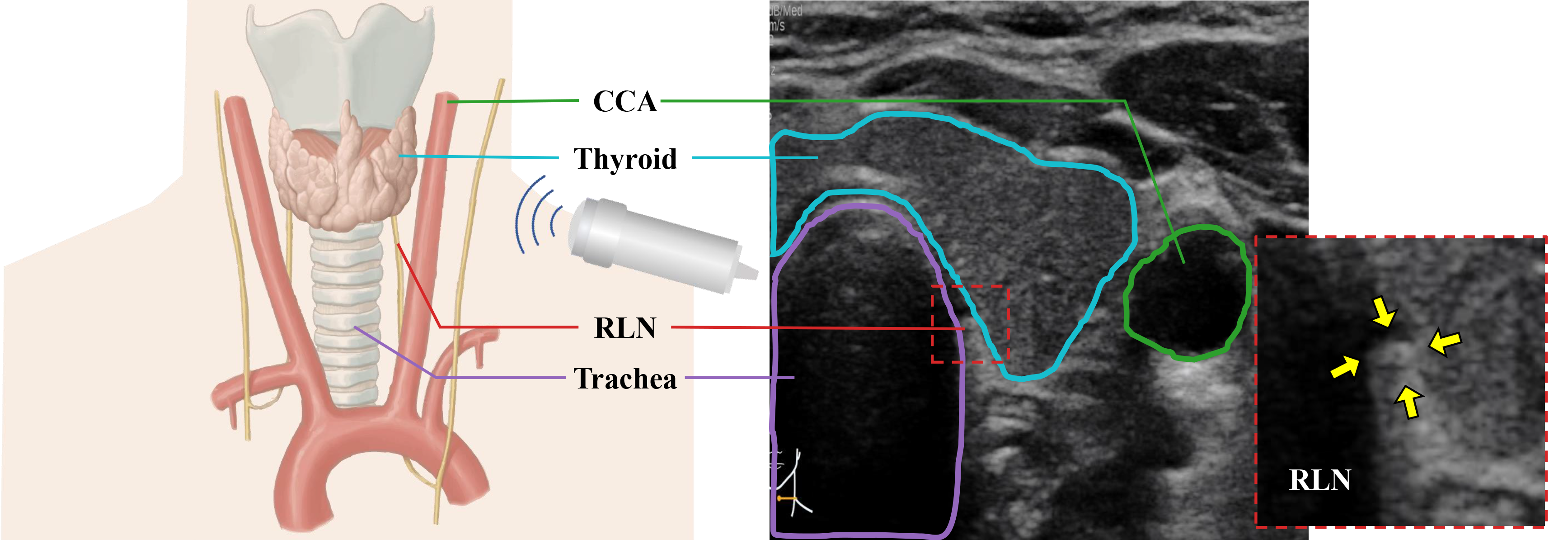}
    \caption{Ultrasound imaging of bilateral RLNs.}
    \label{fig:background}
\end{figure*}

The RLN in the US image is rather tiny relative to organs like the trachea and thyroid (Fig.~\ref{fig:background}). Therefore, automatic detection of the RLN from an US image is a challenging task. Recent studies \cite{MingHuwiHorng2020DeepNerveAN,JuulPAvanBoxtel2021HybridDN,HuisiWu2021RegionawareGC} have demonstrated the feasibility of segmenting large nerves from US images, i.e., the sciatic nerve (diameter ranging from 16 to 20\,mm~\cite{RobertTiel2011NerveIO}) and the median nerve (cross-sectional area ranging from 6.1 to 10.4\,mm$^{2}$~\cite{PMeyer2018TheMN}). These studies propose modifications to basic segmentation models~\cite{ronneberger2015u,KaimingHe2016DeepRL} to improve segmentation accuracy. For example, van Boxtel et al.~\cite{JuulPAvanBoxtel2021HybridDN} investigated the efficacy of a hybrid model on nerve segmentation in US images. Horng et el.~\cite{MingHuwiHorng2020DeepNerveAN} integrated a ConvLSTM block~\cite{BernardinoRomeraParedes2015RecurrentIS} at the bottom layer of the U-Net to capture long-term spatial dependencies. Wu et al.~\cite{HuisiWu2021RegionawareGC} employed multi-size kernels and a pyramid architecture to aggregate features for segmentation. Despite these advances, localization of the RLN remains challenging since it is tiny with mean diameter ranging from 1 to 3\, mm~\cite{BeataWojtczak2018AFA}, significantly smaller than the larger nerves mentioned above. 

In this work, we propose a knowledge-driven framework for RLN localization, mimicking the standard approach surgeons take to identify the RLN according to its surrounding organs.
Our primary contributions are as follows: (1) We propose the first learning-based framework to identify the RLN from a US image for pre-operative assessment of contraindication for robotic thyroidectomy; (2) We introduce Bayesian shape alignment for geometrical constraints, allowing the utilization of spatial prior knowledge in determining an ROI enclosing the RLN; (3) We introduce Locate-Net, a dual-path network that uses both local and global information to refine the localization of the RLN centroid.  

\section{Methods}
\label{method}

\begin{figure*}[t]
	\centering
	\includegraphics[width=\textwidth]{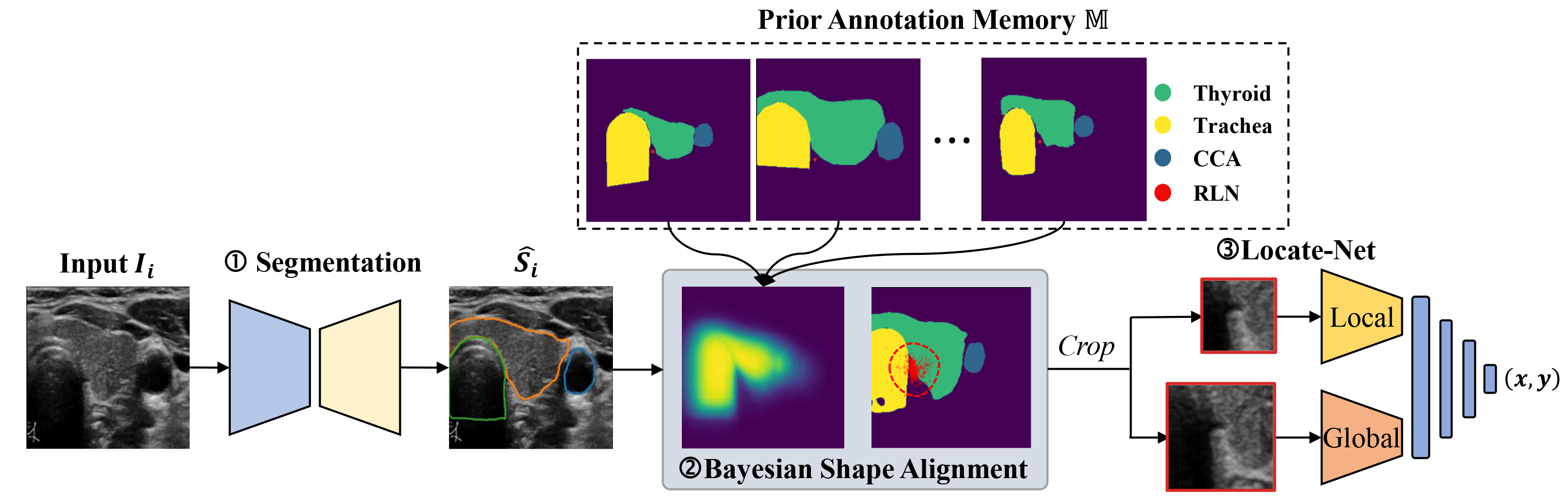}
	\caption{Overview of the proposed framework.}
	\label{fig:framework}
\end{figure*}

Fig.~\ref{fig:framework} illustrates the proposed framework for identifying the RLN from a US image using anatomical prior knowledge. Our framework includes three coarse-to-fine sequential modules: (1) Segmentation module; (2) Bayesian shape alignment (BSA) module; and (3) Locate-Net module. The segmentation module obtains the segmentations $\hat{S}$ of organs surrounding the RLN, including the common carotid arteries (CCA), thyroid, and trachea. These segmentations form posteriors for the BSA module to infer the candidate coordinates of the RLN. Finally, Locate-Net refines the RLN centroid using local details and global contexts based on the patch centered at the inferred candidate coordinates. Details on the three modules are described in the following sections.

\subsection{Bayesian Shape Alignment}
To avoid missing the tiny RLN~\cite{BeataWojtczak2018AFA} in the midst of significantly larger structures, we introduce a method based on anatomical prior knowledge for RLN detection.  Clinically, surgeons recognize the RLN based on its surrounding CCA, thyroid, and trachea~\cite{KaiPunWong2013EndoscopicTA}.  The spatial relationships of these anatomical structures are typically consistent, but not entirely identical, across individuals. Here, we incorporate the spatial relationship into the Bayesian inference with the following mathematical model for a given image $I$: 
\begin{equation}
    \label{eq:mathmodel}
    q(\text{RLN},\text{SO}|I)\propto \\
    p(I|\text{RLN},\text{SO})\times p(\text{RLN}|\text{SO})\times p(\text{SO})
\end{equation}
where $\text{SO}$ is a matrix of pixels of a segmentation image, classifying whether each pixel belongs to the surrounding organs (CCA, trachea, thyroid); $\text{RLN}$ refers to a set of center points, where each point is specified by a location vector $(x,y)$ for the image matrix. $p(\text{SO})$ is the prior distribution of the segmentation maps of the surrounding organs; and $p(\text{RLN}|\text{SO})$ is the likelihood of the RLN centroid given the segmented matrix of its surrounding organs. $p(\text{SO})$ and $p(\text{RLN}|\text{SO})$ depend on the observed cohort and can be taken as prior knowledge. $p(I|\text{RLN},\text{SO})$ represents the joint likelihood for the surrounding organs and the RLN's centroid, and is treated as a constant; and $q(\text{RLN},\text{SO}|I)$ is the likelihood of the RLN centroid and the segmented matrix of its surrounding organs given a particular image $I$.  

In Eq.~\ref{eq:mathmodel}, we aim to predict the RLN's centroid $(x,y)$ given a image $I$ by obtaining the maximum likelihood probability from the priors. The prior distributions for both the RLN centroids and its surrounding segmentation dependent on the given cohort. For each sample $I_{j}$ from training set, the approximate probability $p_{j}(\text{RLN}|\text{SO})$ and $p_{j}(SO)$ attain the maximum value when $I_{j}$ is most similar to the given sample $I$. Based on this, we can infer the likelihood RLN's centroids from the samples with similar surrounding segmentation matrix.

\subsubsection{Prior Distribution for RLN's Surroundings}
\label{subsec:seg}
Segmenting the organs surrounding the RLN is a prerequisite to determining $p(\text{SO})$. Here, the widely adopted segmentation model, U-Net~\cite{ronneberger2015u}, is employed to segment the CCA, thyroid, and trachea from a US image. The segmentation network takes an US image and outputs the corresponding segmentation maps of the three organs. It comprises an encoder and decoder with a skip-connection to forward the feature representations from each stage of the encoder to the corresponding stage in the decoder. The numbers of feature maps in the encoder are 64, 128, 256, 512, 1024, and similarly in the decoder. Each stage in the segmentation network contains two convolution layers followed by instance normalization~\cite{ulyanov2016instance} and ReLU functions~\cite{xu2015empirical}. The training loss function is composed with a cross-entropy loss and a dice similarity coefficient (DSC) loss:
\begin{equation}
    \label{eq:segmentaion}
    \mathcal{L}_{\text{seg}} = \mathcal{L}_{\text{ce}}(\hat{S},M) + \mathcal{L}_{\text{dsc}}(\hat{S},M)
\end{equation}
where $\mathcal{L}_{\text{ce}}$ and $\mathcal{L}_{\text{dsc}}$ refer to cross-entropy loss and DSC loss, respectively.

\subsubsection{Alignment-based Likelihood}
Employing the prior sample $I_j$ that is similar in the surrounding masks with the given image $I_i$ to derive the RLN's centroid can derive higher likelihood $p(\text{RLN}|\text{SO})$. However, affected by the probe scanning angles $\theta$, the observations belong to different angle distributions and can not be directly used to construct the priors of RLN's surrounding segmentation, hence, we embed a pre-alignment module to eliminate the influence of $\theta$. The detailed implementation of the proposed alignment-based likelihood infer for the RLN is described in Algorithm~\ref{alg:align}.

\begin{algorithm}[t]
    \caption{Bayesian shape alignment during inference}
    \label{alg:align}
    \begin{algorithmic}
        \Require The predicted segmentation mask $\hat{S}_{i}$ for a inference sample, a set of mask labels $\mathbb{M}=\{M_j,j\in\mathbb{N}^{*}\}$ and RLN labels $\mathbb{C}=\{c_j,j\in\mathbb{N}^{*}\}$ for the training samples
        \Ensure Candidate coordinate $\mathcal{C}_i$ of RLN for $\hat{S}_{i}$
        \State $D[*]=\{d_j,j\in\mathbb{N}^{*}\}$
        \State $C[*]=\{p_j,j\in\mathbb{N}^{*}\}$
        \For{corresponding label $\{M_j$,$c_j\}$ in $\{\mathbb{M},\mathbb{C}\}$}
        \State $\phi_{j} \gets {\rm Affine}(M_j,\hat{S}_{i})$ \Comment{shape analysis}
        \State $D[j] \gets {\rm Dice}(M_{j}\circ\phi_{j},\hat{S}_{i})$ \Comment{calculate dice metric}
        \State $C[j] \gets c_{j}\circ\phi_{j}$ \Comment{transform centroid of RLN}
        \EndFor
        \State $index \gets {\rm Rank}(D[*])$  \Comment{rank dice in the descending order}
        \State $C_{sort} \gets {\rm Sort}(C[*],index)$  \Comment{sort coordinates with dice rank}
        \State $\mathcal{C}_{i} \gets {\rm AverageTopK}(C_{sort})$  \Comment{average the Top-$k$ candidate coordinates}
    \end{algorithmic}
\end{algorithm}

Algorithm~\ref{alg:align} estimates the likelihood RLN's centroid by searching the similar samples from the priors. We first align the priors to the common angle distribution. Taking each predicted mask $\hat{S}_{i}$ as the target segmentation and every transversed annotation $M_j$ in the training dataset $\mathbb{M}$ as the moving segmentation, the affine transformation matrix $\phi_{j}$ with 6 degrees of freedom (DOF) can be conducted with the mutual information as the similarity metric. After such alignment, the Dice ratio (DSC) between the predicted mask $\hat{S}_{i}$ and each affined segmentation $M_{j} \circ \phi_{j}$ in the training dataset is calculated as the shape similarity score. More similar in the surrounding segmentation correspond to higher likelihood estimation for the entire image and the RLN's centroid. Subsequently, we descend the priors depending on the DSC, and average the RLNs' candidates with the Top-$k$ DSCs to infer the position of RLN for the given image $I_{i}$.

\subsection{Locate-Net}
The BSA module is followed by a dual-path neural network to further refine the centroid of the RLN. Based on the candidate center coordinates of the RLN determined by the BSA module, a global patch with the size of $64\times64$ and a local patch with the size of $24\times24$ are cropped from the US image and jointly fed to the refinement network to provide global and local information. The features from the global patch contain global semantics to capture the potential outlier unobserved in the priors. The features from the local patch provide details to help refine the centroid. 

In the refinement network, features from the local and global patches are separately extracted with two weight-shared sub-networks. Each sub-network contains three convolutional blocks. After each convolution block, the feature maps are down-sampled with a factor of 2 via the max-pooling layer. The detailed composition of each convolutional block is the same with the encoder in the segmentation network (seen in Section~\ref{subsec:seg}). The outputs of the two sub-networks are then re-scaled to the same size with a adaptive pyramid pooling layer~\cite{zhao2017pyramid}, and followed with a concatenation layer and two convolution blocks. The top of the refinement network is three fully connected layers with the hidden units of 512, 64, 2, so that to predict the refined centroid of RLN. The regression loss function for training the refinement network is defined as:
\begin{equation}
    \label{eq:regression}
    \mathcal{L}_{\text{reg}} = \sum_i\mathcal{L}_{s1}(\hat{c}_i-c_i),
\end{equation}
where $\mathcal{L}_{s1}$ is the smoothed $L_1$ loss with $\beta$ of 1.0 indexed for the difference of image pixel $(\Delta x,\Delta y)$, $\hat{c}$ and $c$ denote as the predicted RLN centroid and the ground truth, respectively.
\begin{equation}
    \mathcal{L}_{s1}(\Delta x,\Delta y) = 
            \begin{cases}
                \frac{1}{2\beta}({\Delta x}^2 + {\Delta y}^2)  &  {\left|\Delta x\right|+\left|\Delta y\right|<2\beta},\\
                \left|\Delta x\right|+ \left|\Delta y\right|- \beta &  \textrm{others}.
            \end{cases}
\end{equation}

\section{Experiments}
\subsection{Dataset and Evaluation Metrics}
2D ultrasound images were collected with an Aixplorer color Doppler ultrasound device (Hologic Supersonic imagine, AIX en Provence), equipped with a linear array probe with a frequency of $4$--$15$\,MHz. A total of 465 patients diagnosed with thyroid cancer by preoperative biopsy and enrolled for thyroidectomy participated in this study. Each patient has both left and right scans of the RLN.  Each scan contains a variable number of qualified US frames, ranging from 1 to 4. 325, 46 and 94 subjects were randomly selected for training, validation, and testing, respectively. All images were resampled to a common size of $256\times256$. Manual annotation was contented by three clinical experts and passed through strict quality control from a senior expert.

Performance was quantified using the absolute distance error and the hit rate between the predicted RLN centroid and the ground truth:
\begin{equation}
    \label{eq:dis}
    \texttt{Dis}(\hat{c}, c) = \|\hat{c}-c\|_{1}
\end{equation}
\begin{equation}
    \label{eq:hit}
        \texttt{Hit}(\hat{c}, c) = 
        \begin{cases}
        1 & \hat{c} \subseteq N_{c},\\
        0 & \textrm{others},\\
        \end{cases}
\end{equation}
where $\hat{c}$ and $c$ denote as the predicted RLN centroid and the ground truth, respectively. $N_c$ indicates the neighborhood of $c$ with radius $r_{\theta}$, which is set to be 15 pixels in this work. Lower distance error and higher hit rate correspond to better identification performance.

\subsection{Implementation Details}
We implemented our method using PyTorch on the Google Colab platform with an NVIDIA Tesla P100 GPU. We trained the segmentation network using the Adam optimizer with an initial learning rate of $3\times 10^{-4}$ and a batch size of 16 for 100 epochs, taking about 3 hours. The learning rate was decayed every epoch with a factor of 0.9. The affine matrix in Bayesian shape alignment was initially computed with the center of mass of the masks and iteratively refined with the mutual information metric. For the training of the dual-path refinement network, the learning rate and batch size were set to $1\times10^{-3}$ and 16, respectively. The code for the techniques presented in this study can be found at: https://github.com/wulalago/RLNLocalization

\subsection{Comparison Baselines}
\begin{figure*}[t]
    \centering
    \includegraphics[width=0.8\textwidth]{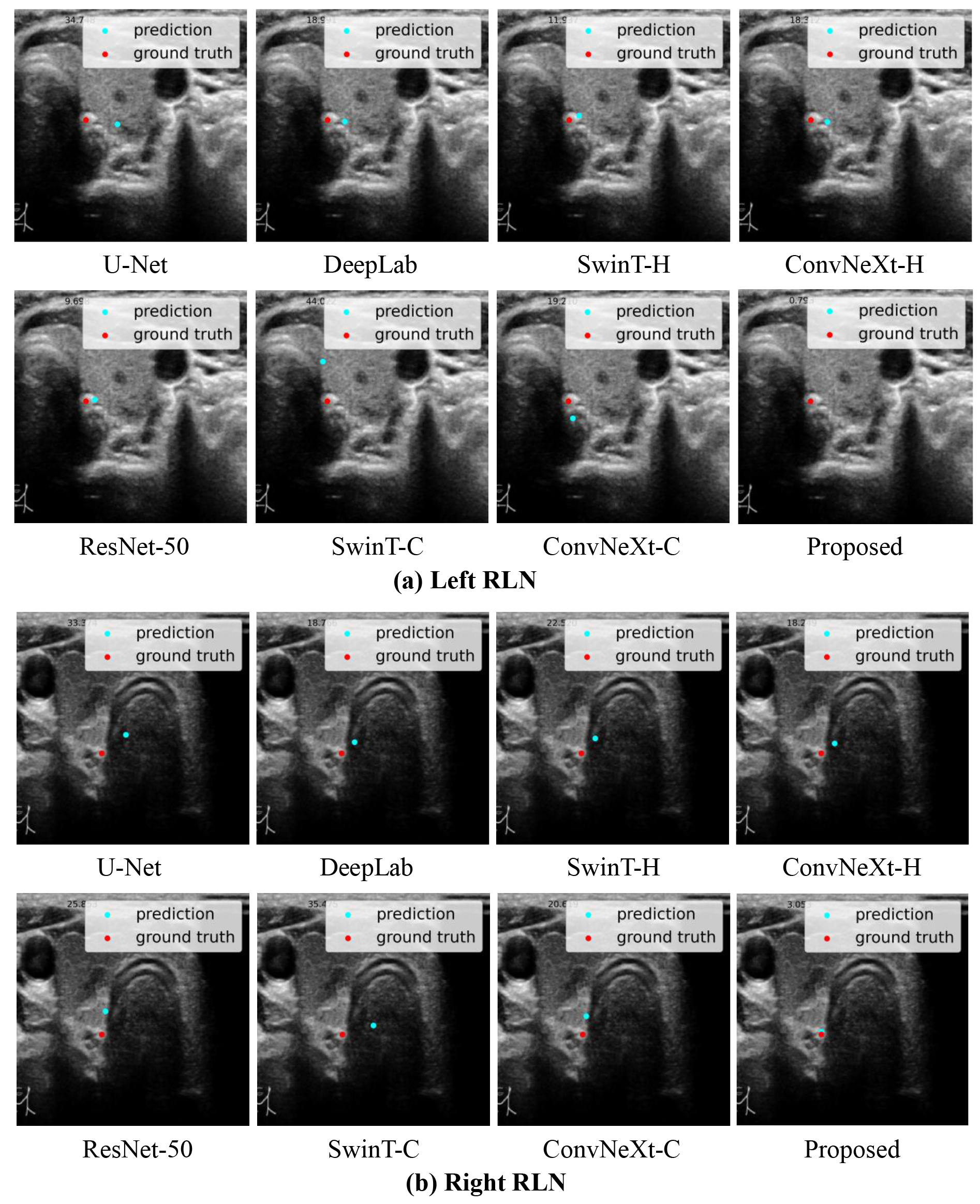}
    \caption{Example results from competing methods.}
    \label{fig:baselines}
\end{figure*}

We compared our method with coordinate and heatmap regression methods. Compared heatmap regression methods include U-Net~\cite{ronneberger2015u}, DeepLab~\cite{BowenCheng2020PanopticDeepLabAS}, SwinT-H~\cite{ZeLiu2021SwinTH}, and ConvNeXt-H~\cite{liu2022convnet}. Coordinate regression methods include ResNet-50~\cite{KaimingHe2016DeepRL}, SwinT-C~\cite{ZeLiu2021SwinTH}, and ConvNeXt-C~\cite{liu2022convnet}. The optimal hyper-parameters of all the baseline methods are obtained based on the grid search strategy. Fig.~\ref{fig:baselines} shows example cases of the bilateral RLNs given by the the baseline methods. The red and cyan circles mark the annotated ground truth and the predicted centroid of the RLN, respectively. Our method predicts the centroids of the bilateral RLNs with higher accuracy than the baseline methods. Table~\ref{tab:baselines} reports the statistics of the results given by the methods on the testing dataset. The proposed method achieves the lowest distance error with the highest hit rate.

\begin{table}[t]
    \centering
    \renewcommand{\arraystretch}{1.3}
    \caption{Statistics of competing methods for the testing dataset.\label{tab:baselines}}
    \begin{tabular}{llcccccc}
        \hline \hline
        & \multirow{2}{*}{Methods} & \multicolumn{2}{c}{Left RLN} & \multicolumn{2}{c}{Right RLN} \\
        \cline{3-4} \cline{5-6}
        & & Distance ($pix$) & Hit Rate ($\%$) & Distance ($pix$) & Hit Rate ($\%$)  \\
        \hline
        \multirow{4}{5em}{Coord-based}
            & ResNet-50~\cite{KaimingHe2016DeepRL} & 10.9 $\pm$ 9.7 & 77.5 & 12.3 $\pm$ 8.4 & 70.0\\
            & SwinT-C~\cite{ZeLiu2021SwinTH} & 19.7 $\pm$ 11.8 & 42.3 & 14.4$\pm$ 9.6 & 59.4 \\
            & ConvNeXt-C~\cite{liu2022convnet} & 16.5 $\pm$ 8.2 & 47.3 & 14.0 $\pm$ 9.5 & 62.5\\
        \hline
        \multirow{4}{5em}{Heatmap-based}
            & U-Net~\cite{ronneberger2015u} & 29.3 $\pm$ 12.8 & 11.5 & 20.9 $\pm$ 10.5 & 31.9\\
            & DeepLab~\cite{BowenCheng2020PanopticDeepLabAS} & 17.5 $\pm$ 8.0 & 41.2 & 11.9 $\pm$ 7.1 & 71.3\\
            & SwinT-H~\cite{ZeLiu2021SwinTH} & 22.7 $\pm$ 13.6 & 30.8 & 20.2 $\pm$ 10.6 & 35.6\\
            & ConvNeXt-H~\cite{liu2022convnet} & 12.7 $\pm$ 12.1 & 73.1 & 13.1 $\pm$ 8.7 & 64.4\\
        \hline
        & Proposed & \textbf{3.49$\pm$7.53} & \textbf{95.6} & \textbf{4.55 $\pm$7.61} & \textbf{92.5}\\
        \hline \hline
    \end{tabular}
\end{table}

\begin{table}[t]
    \centering
        \renewcommand{\arraystretch}{1.3}
    \caption{Statistics for different settings based on the testing dataset.\label{tab:ablation}}
    \begin{tabular}{llcccccc}
    \hline \hline
    & \multirow{2}{*}{Methods} & \multicolumn{2}{c}{Left RLN} & \multicolumn{2}{c}{Right RLN} \\
    \cline{3-4} \cline{5-6}
    & & Distance ($pix$) & Hit Rate ($\%$) & Distance ($pix$) & Hit Rate ($\%$)  \\
    \hline
    & Initialization & 7.52$\pm$7.65 & 89.0 & 9.70$\pm$7.03 & 84.9\\
    & + Local information & 4.45$\pm$8.79 & 91.8 &5.47$\pm$7.90 & 88.7\\
    & + Global context & 3.58$\pm$7.51 & 94.5 &4.42$\pm$7.48 & 91.8\\
    & + Local \& global features & \textbf{3.49$\pm$7.53} &  \textbf{95.6} & 4.55 $\pm$7.61 & \textbf{92.5}\\
    \hline \hline
    \end{tabular}
\end{table}

\subsection{Ablation Study}
We compared three types of centroid refinement methodes, including (1) refinement with local information; (2) refinement with global contextual information; and (3) refinement with both local and global features. Table~\ref{tab:ablation} reports the distance errors and hit rates for these settings based on the proposed method. Refinement with local and global information yields the lowest distance error with the highest hit rate.

\section{Conclusion}
Inspired by the way surgeons to recognize RLN, we developed a prior knowledge driven framework to automatically identify the tiny RLN from US images. In the proposed pipeline, we first segment the large organs surrounding RLN as the conditional prior, and then using the Bayesian shape alignment model to determine the candidate coordinate close to RLN. Then following the Locate-Net to refine the centriod of RLN with multi-scales patches to extract the local information and global context around the RLN. Leveraging the spatial relationship between RLN and its surrounding organs as the prior constraint, our model can avoid the tiny RLN being submerged the background. From Tabel~\ref{tab:baselines} and Tabel~\ref{tab:ablation}, we can conclude that, any combination of our framework achieves the superiority in the distance error and hit rate as compared to the recent coordinate or heatmap regression models. 
\subsubsection{Acknowledgement}
This work was supported by the National Natural Science Foundation of China under Grant No. 62101365 and the startup foundation of Nanjing University of Information Science and Technology. Haoran Dou was funded by the Royal Academy of Engineering Chair in Emerging Technologies Scheme (CiET1819/19). Luyi Han was funded by Chinese Scholarship Council (CSC) scholarship. 
Alejandro F. Frangi was funded by the Royal Academy of Engineering (INSILEX CiET1819/19), Engineering and Physical Sciences Research Council (EPSRC) programs TUSCA EP/V04799X/1, and the Royal Society Exchange Programme CROSSLINK  IES$\backslash$NSFC$\backslash$201380. 
Jun Xu was funded by the National Natural Science Foundation of China (Nos. U1809205, 62171230, 92159301, 61771249, 91959207, 81871352).
%
%
\bibliographystyle{splncs04}
\bibliography{ref}

\end{document}